\documentclass[%
 reprint,
superscriptaddress,
 amsmath,amssymb,
 aps,
]{revtex4-2}

\usepackage{graphicx}
\usepackage{dcolumn}
\usepackage{bm}
\usepackage{hyperref}

\usepackage{subfigure}
\usepackage{amsmath}

\usepackage{color}

\begin{document}

\title{Stochastic behaviour of an interface-based memristive device}

\author{Sahitya Yarragolla}
\affiliation{Chair of Applied Electrodynamics and Plasma Technology, Ruhr University Bochum, Germany}

\author{Torben Hemke}
\affiliation{Chair of Applied Electrodynamics and Plasma Technology, Ruhr University Bochum, Germany}

\author{Jan Trieschmann}
\affiliation{Electrodynamics and Physical Electronics Group, Brandenburg University of Technology Cottbus-Senftenberg, 03046 Cottbus Germany}

\author{Finn Zahari}
\affiliation{Nanoelectronics, Faculty of Engineering, Christian Albrechts University of Kiel, 24143 Kiel, Germany}

\author{Hermann Kohlstedt}
\affiliation{Nanoelectronics, Faculty of Engineering, Christian Albrechts University of Kiel, 24143 Kiel, Germany}

\author{Thomas Mussenbrock}
\affiliation{Chair of Applied Electrodynamics and Plasma Technology, Ruhr University Bochum, Germany}

\date{\today}

\begin{abstract}
A large number of simulation models have been proposed over the years to mimic the electrical behaviour of memristive devices. The models are based either on sophisticated mathematical formulations that do not account for physical and chemical processes responsible for the actual switching dynamics or on multi-physical spatially resolved approaches that include the inherent stochastic behaviour of real-world memristive devices but are computationally very expensive. In contrast to the available models, we present a computationally inexpensive and robust spatially 1D model for simulating interface-type memristive devices. The model efficiently incorporates the stochastic behaviour observed in experiments and can be easily transferred to circuit simulation frameworks. The ion transport, responsible for the resistive switching behaviour, is modelled using the kinetic Cloud-In-a-Cell scheme. The calculated current-voltage characteristics obtained using the proposed model show excellent agreement with the experimental findings.
\end{abstract}

\maketitle

\section{\label{sec:sec1.0} Introduction}

For over 50 years, Moore's law has been the guiding principle of the electronics world. However, recent times have witnessed a breakdown of this law, with modern devices driving the future of computing beyond Moore's law \cite{Moore1965}. Memristive devices (or memristors) are a kind of `More-than-Moore' devices that have become a hot topic of interest for researchers in the field of solid-state physics, materials science, electronics and computer engineering \cite{Waldrop2016,IRDS2021}. In its generic form, a memristive device is a two-terminal device with an active (often electrically insulating) layer sandwiched between two metal electrodes. In simple terms, it is nothing but a non-linear resistor with memory, in the sense that the Ohmic resistance depends on the history of the current that has flown through it driven by an applied voltage \cite{Chua2011,Waser2021}. The switching from a low resistance state (LRS) to a high resistance state (HRS) and vice versa is initiated once the driving voltage reaches a specific value. With features such as low power consumption, passivity, and scalability down to the nanometer range, these devices have emerged out as potential candidates in the field of hardware security~\cite{Du2021,John2021}, non-volatile memory applications (e.g., resistive random access memories, RRAMs)~\cite{Jung2021,Xu2015,Chen2016}, and neuromorphic systems~\cite{Hansen2017,Yang2013,Zidan2018,Lin2020}. 

State-of-the-art memristive devices based on field-driven nano-ionic transport can be either interface-type or filamentary-type. The latter relies on the formation and dissolution of conducting filaments. These conducting filaments may consist of metal atoms or cations in so-called electrochemical metallization (ECM) cells~\cite{Valov_2011,Menzel2015,Dirkmann2015,Dirkmann2017} or oxygen vacancies in, e.g., hafnium oxide valence change mechanism (VCM) cells~\cite{Dirkmann2018}. The switching behaviour of such filamentary devices is often digital and fast, and the states are stable. Therefore, the applications are predominantly in the field of RRAMs. In contrast, interface type devices are often driven by the movement of ions or charged defects within the active layer that modify the electrical characteristics of the metal/insulator interfaces (Schottky contacts or tunneling contacts), which act as boundaries for the active layer~\cite{Hansen2015,You2014,Park2021}. The switching dynamics is therefore analogue. The states are less stable, so the application of these devices is more in the field of neuro-inspired electronics. An important feature common in almost all nano-ionic memristive devices is that they show intrinsic stochastic behaviour due to the involved ion movement.

A number of different models have been proposed to explain the operation and switching dynamics of the devices on the experimental time scale of seconds to days. The actual switching mechanism on the atomic time scale is seldom addressed \cite{Onofrio2015}. Although the models are able to mimic the electrical behaviour of the devices, they all have certain shortcomings. Concentrated models are based on complex mathematical formulations of the switching dynamics~\cite{Solan2017,Ambrogio2014,Hardtdegen2018}. Despite being fast, the physical and chemical processes responsible for the actual switching dynamics are immensely neglected, and thus, they mostly show deterministic behaviour. Besides, more elaborate stochastic models are based on multi-physical 2D or 3D approaches that include more or less realistic physics. For example, 3D kinetic Monte-Carlo (kMC) based simulation models produce characteristics similar to real-world memristive devices, successfully capturing their intrinsic stochastic behaviour~\cite{Dirkmann2016,Dirkmann2018,Abbaspour2016,Marcel2017}. Nevertheless, these models are computationally expensive, and therefore, it is not feasible to couple them with circuit simulators to address memristive systems.  The drawbacks of the models mentioned above form the prime motivation for our work.   

In the framework of the present work, a spatially one-dimensional physics-inspired model that considers the stochastic nature of resistive switching is proposed. It is focused on the random behaviour observed in a real-world interface type device, showing a comparable cycle-to-cycle and device-to-device variation while conserving the physics. The model is based on Newton's laws of motion for the ion motion in the active layer combined with the kinetic Cloud-In-a-Cell (CIC) scheme~\cite{Laux1996} (also known as the Particle-In-Cell scheme, which is often used in plasma physics) to couple the ion transport mechanism with the electric field within the device. A feature of this approach is that it is deterministic. However, the stochastic behaviour comes into play by randomly perturbed charge positions within the CIC scheme. It is important to notice that the proposed model is computationally less expensive than the usual kMC-based models. Therefore, it is compatible with SPICE-level or Verilog-A circuit simulations.

The proposed stochastic model and its effectiveness are discussed in detail, taking the double barrier memristive device (DBMD)~\cite{Hansen2015} as the subject device. To understand the proposed model, we start with the introduction of the DBMD device in Section~\ref{sec:sec2.0}. The description of the simulation model is then given in Section~\ref{sec:sec3.0}, including the implementation of the numerical scheme, stochasticity modelling and the current transport mechanisms. Finally, the results are discussed in Section~\ref{sec:sec4.0}, and conclusions are drawn in Section~\ref{sec:sec5.0}.

\section{Device configuration}
\label{sec:sec2.0} 

\begin{figure}[!t]
\includegraphics[width=1\columnwidth]{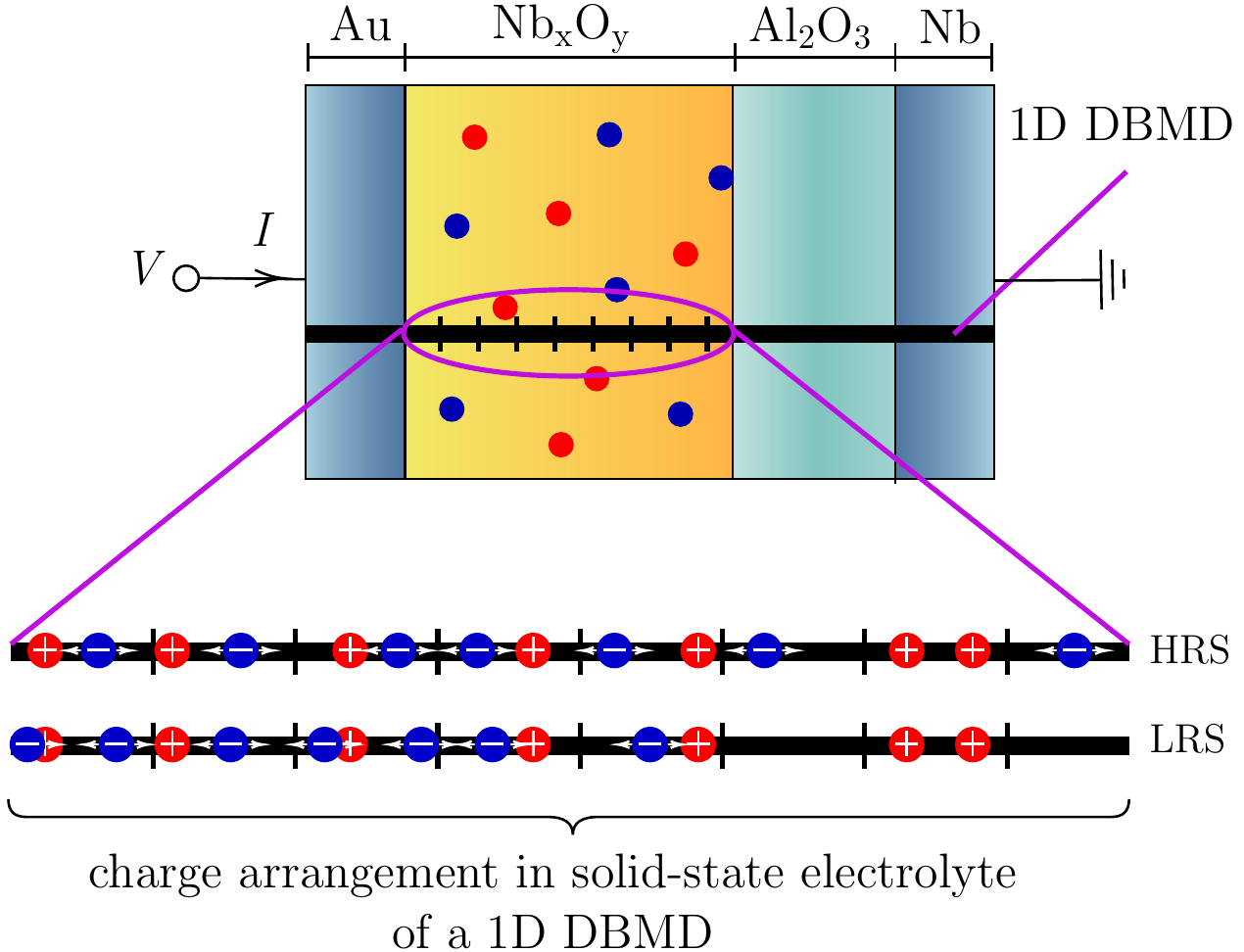}
\caption{An interface type double barrier memristive device with an $\rm {Au/Nb_{x}O_{y}}$ Schottky contact, $\rm Nb_{x}O_{y}$ solid-state electrolyte and $\rm Al_{2}O_{3}$ tunnel barrier \cite{Dirkmann2016}. A 1D DBMD is shown as a thick black line with fixed ions (red circles) and mobile ions (blue circles). }
\label{fig:2-1} 
\end{figure}

The interface type double barrier memristive device, shown in Fig.~\ref{fig:2-1}, was introduced by Hansen et al.~\cite{Hansen2015}. The device mainly consists of three interfaces in the order $\rm Au/Nb_{x}O_{y}/Al_{2}O_{3}/Nb$. The $\rm {Au/Nb_{x}O_{y}}$ interface acts as a Schottky contact, which is assumed to be inert to prevent oxidation. The $\rm Al_{2}O_{3}$ tunnel barrier is an electrically high-quality barrier that allows for elastic electron tunnelling. The $\rm Nb_{x}O_{y}$ layer in the centre represents a solid-state electrolyte that acts both as an electronic and ionic conductor. The mobile negatively charged oxygen ions (blue circles) in this layer are primarily responsible for changing the resistance of the device, while the stationary positively charged oxygen ions (red circles) maintain charge neutrality. Under the influence of an externally applied electric field, the negatively charged oxygen ions drift within the $\rm Nb_{x}O_{y}$ layer (the charge arrangement in HRS and LRS is indicated in  Fig.~\ref{fig:2-1}). This oxygen ion drift eventually influences the interface properties of both the Schottky-contact and the tunnel barrier, leading to an overall change in device resistance. While the field-induced ion movement can model the switching mechanism very well, as it is decribed in detail in~\cite{Hansen2015,Dirkmann2016}, the charging and discharging of interface traps cannot be completely disregarded as switching mechanism~\cite{Hansen2015}. It should be noted that the device stack shown in~\cite{Hansen2015,Dirkmann2016} differs slightly from the here presented one. It has been shown by electron energy loss spectroscopy, that the few nanometer thin $\rm Al$ bottom electrode is fully oxidized and even the $\rm Nb$ layer underneath is slightly oxidized to a sub-stoichiometric $\rm {NbO_{z}}$ at the interface~\cite{Strobel2017}. However, this does neither influence the model shown in~\cite{Hansen2015,Dirkmann2016} nor the 1D model presented here, because the $\rm {NbO_{z}}$ is believed to be a metallic resistor and tunneling through the alumina is still reasonable~\cite{Strobel2017}. The device parameters used in the 1D model in this work are extracted from~\cite{Hansen2015} and are summarized in Table~\ref{table:3_1}.

\begin{table}[b]
\caption{\label{table:3_1}%
Details of the simulation parameters~\cite{Hansen2015}
}
\begin{ruledtabular}
\begin{tabular}{l l l}
\textrm{Physical quantity}&
\textrm{Symbol}&
\textrm{Value}\\
\colrule
Temperature & $T$ & 300K \\
Lattice constant & $d$ & $2.5 \times 10^{-10}\, {\rm m}$\\
Device area & $A_{\rm d}$ & $625\, \mu{\rm {m}^{2}}$\\
Relative permittivity ($\rm Nb_{x}O_{y}$) & $\varepsilon_{r} $ & 42.0\\
Activation energy & ${\cal E}_{\rm A}$ & 0.76\\
Conductivity ($\rm Nb_{x}O_{y}$) & $\sigma$ & $1.0 \times
 10^{-4}\, \Omega {\rm m}$\\
 Length of SE ($\rm Nb_{x}O_{y}$) & $l_{\rm SE}$ & $2.5 \times 10^{-9}\, {\rm m}$\\
Defect density & $\rho$ & $5\times 10^{20} {\rm cm^{-3}} $\\
Tunnel barrier width & $d_{\rm 0}$& $1.1 \times 10^{-9}\, {\rm m}$\\
Tunnel barrier height & $\Phi^{\rm t}$ & $3.2$ eV\\
Schottky barrier height & $\Phi^{\rm b}$ & $0.98$\, eV\\
Schottky barrier ideality factor & $n$ & 4.2\\
\end{tabular}
\end{ruledtabular}
\end{table}

\section{Model description}
\label{sec:sec3.0}

The equivalent circuit of DBMD is used as the starting point for the electrical representation of the compact model. In contrast to the equivalent circuit of DBMD by Hansen et al.~\cite{Hansen2015}, a simplified version of the equivalent circuit is adopted. The equivalent circuit diagram of the DBMD is shown in Fig.~\ref{fig:2-2}. The Schottky contact is represented as a Schottky diode and the tunnel barrier as a voltage-controlled current source. The ion distribution in the solid-state electrolyte is expressed as a resistance ($R_{\rm SE}$). By applying Kirchhoff's voltage law (KVL) and Kirchhoff's current law (KCL) to the equivalent circuit, we obtain the following equations,

\begin{equation}
    V_{\rm Device} = V_{\rm SC} + V_{\rm SE}+V_{\rm TB}, 
    \label{Eq:2.1}
\end{equation}

\begin{equation}
    I_{\rm SC} = I_{\rm SE} = I_{\rm TB} = I.
    \label{Eq:2.2}
\end{equation}

\noindent$V_{\rm Device}, V_{\rm SC}, V_{\rm SE}, V_{\rm TB}$ are the voltage drops across the device, Schottky contact, solid-state electrolyte and tunnel barrier, respectively. Similarly, $I_{\rm SC}, I_{\rm SE}, I_{\rm TB}$ are the currents across the Schottky barrier, solid-state electrolyte and tunnel barrier, respectively. Based on the equivalent circuit diagram and Eqs.~\eqref{Eq:2.1}-\eqref{Eq:2.2}, the ion transport and the current mechanisms are now implemented and discussed in the following subsections.

\subsection{CIC inspired ion transport}
\label{sec:sec-3.1.1}
As mentioned in Section~\ref{sec:sec2.0}, the overall change in device resistance is mainly due to ion transport. Therefore, it is necessary to choose a suitable transport mechanism that produces characteristics equivalent to a physical device. Unlike the available ion transport models, such as the kMC model by Dirkmann et al.~\cite{Dirkmann2016}, a much simpler transport method inspired by the Particle-In--Cell (PIC) scheme~\cite{Hockney1988,Birdsal2005} is employed in the present work. This method is well known as the Cloud-In-a-Cell (CIC) scheme in the semiconductor community~\cite{Laux1996} .

\begin{figure}[!t]
\includegraphics[width=1\columnwidth]{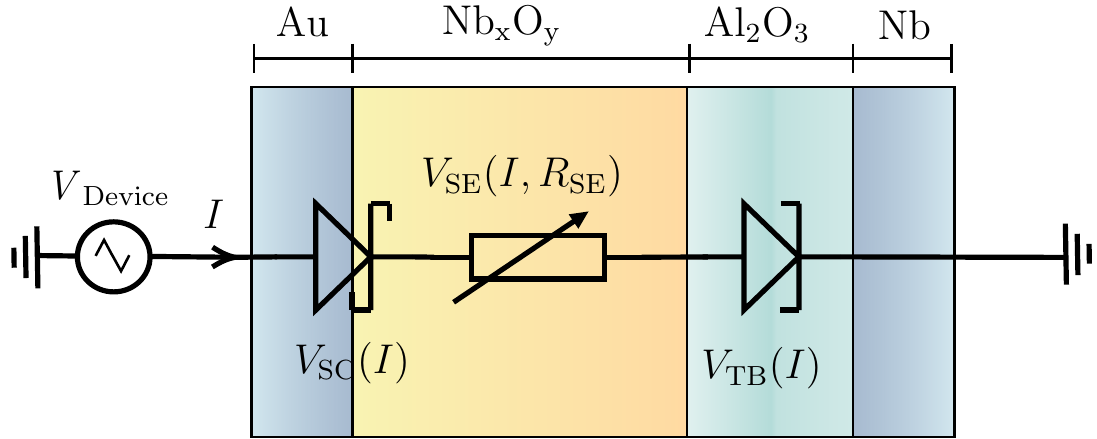}
\caption{The equivalent circuit model of a DBMD with a Schottky diode, solid-state electrolyte as an Ohmic region and the tunnel barrier as a voltage-controlled current source.}
\label{fig:2-2} 
\end{figure}

The computational domain consists of a solid-state electrolyte ($l_{\rm SE}$) modelled on a 1D Eulerian grid. Initially, an equal number of mobile and fixed ions are randomly distributed within the solid-state electrolyte considering a Lagrangian grid. The spatial arrangement of ions in this form is regarded as the HRS depicted in Fig.~\ref{fig:2-1}. For simplicity, from here, we consider two loops, (a) the inner loop for solving Newton's laws and (b) the outer loop for ion transport.\\

\textit{Inner loop:} A basic mathematical optimisation is performed to obtain $V_{\rm TB}$ by minimising the error between $I_{\rm TB}$ and $I_{\rm SC}$. For this, the initial value of $V_{\rm TB}$ is assumed to be $V_{\rm Device}$. Then, the other circuit parameters (mentioned in Fig.~\ref{fig:2-2}) are calculated and reiterated until Eq.~\eqref{Eq:2.2} is satisfied. At the end of each loop, $V_{\rm TB}$ is modified according to the relative error~($\xi 
$) between $I_{\rm TB}$ and $I_{\rm SC}$. Finally, using the voltage drops, the potentials at the $\rm Au/Nb_{x}O_{y}$ interface and $\rm Nb_{x}O_{y}/Al_{2}O_{3}$ interface are respectively given by

\begin{equation}
    \phi_{\rm Au/Nb_{x}O_{y}} = V_{\rm Device} - V_{\rm SC},
    \label{Eq:3.1.1}
\end{equation}
and
\begin{equation}
    \phi_{\rm Nb_{x}O_{y}/Al_{2}O_{3}} = V_{\rm TB}.
    \label{Eq:3.1.2}
\end{equation}

\textit{Outer loop:} Followed by the inner loop, the electric potential is calculated using the 1D Poisson's equation,

\begin{equation}
    \frac{d}{dx} \left ( \varepsilon \frac{d\phi}{dx}\right ) = -\rho,
    \label{Eq:3.1.3}
\end{equation}

\noindent where $\varepsilon $ is the permittivity of $\rm Nb_{x}O_{y}$ and $\rho$ is the charge density. The potential equation is subjected to Dirichlet boundary conditions at the simulation domain boundaries, given by Eqs.~\eqref{Eq:3.1.1} and \eqref{Eq:3.1.2}. The electric field is then obtained from the electric potential by 

\begin{equation}
E = -\frac{d\phi}{dx}.
\label{Eq:3.1.4}
\end{equation}

\noindent This electric field $E$ is used to push the mobile ions with a certain drift velocity assuming the Lagrangian grid. The spatial arrangement of drifted ions in LRS is illustrated in Fig.~\ref{fig:2-1}. In solid-state physics, ion transport on an atomic level is characterised by jump attempts over a potential barrier. Based on this theory, the drift velocity of an ion is given by~\cite{bruce_1994, Meyer2008}

\begin{equation}
    v_{\rm D} = d{p}({\cal E}_{\rm A})\left ( {\rm exp}\left \{ \frac{\left | z \right |edE}{2k_{\rm B}T}
 \right \} - {\rm exp}\left \{ -\frac{\left | z \right |edE}{2k_{\rm B}T}
 \right \}\right ),
 \label{Eq:3.1.5}
\end{equation}

\noindent where $d$ is the jump distance (lattice constant), $z$ is the charge number of the ion, $k_{B}$ is the Boltzmann constant, $T$ is the temperature, and $e$ is the elementary charge. The transition probability, ${p}({\cal E}_{\rm A})$, for an ion to jump to the neighbouring site is

\begin{equation}
    {p}({\cal E}_{\rm A}) = N(1-n_{\rm c})f\nu_{0}  {\rm exp}\left ( -\frac{{\cal E}_{\rm A}}{k_{\rm B}T} \right ),
    \label{Eq:3.1.6}
\end{equation}

\noindent where $n_{\rm c}=0.5\,{\rm exp}\left ( \frac{-\Delta H_{\rm g}}{2k_{\rm B}T}\right )$ is the probability that a site is occupied, then  $(1-n_{\rm c})$ gives the probability that the next neighbour site is empty. $\nu_{0}$ is the phonon frequency, $\Delta H_{\rm g}$ is the intrinsic energy gap, ${\cal E}_{\rm A}$ is the activation energy, $N$ is the number of these empty sites, and $f$ is a geometrical factor of order unity. For a 1D transport, the term $N(1-n_{\rm c})f$ is approximately equal to unity. It must be taken care that the Lagrangian grid is only assumed for distributing the ions and calculating drift velocity, and the rest of the simulation is performed on an Eulerian grid.

As further detailed in the subsequent sections, the mobile ion movement consistently changes the internal state of the device. The average relative distance between the current position of mobile ions and their inertial position determines the internal state in the proposed model. The internal state is given by

\begin{equation}
    q(t) = \frac{\bar{d}(t)-\bar{d}_{\rm r}}{\bar{d}_{\rm r}},
    \label{Eq:3.1.7}
\end{equation}

\noindent with

\begin{equation}
    \bar{d} = \frac{\sum_{i=1}^{N_{\rm ions}}\left ( \bar{x}_{\rm i}-\bar{x}_{\rm SC}\right )}{N_{\rm ions}}.
\end{equation}

\begin{figure}[!t]
\includegraphics[width=0.47\textwidth]{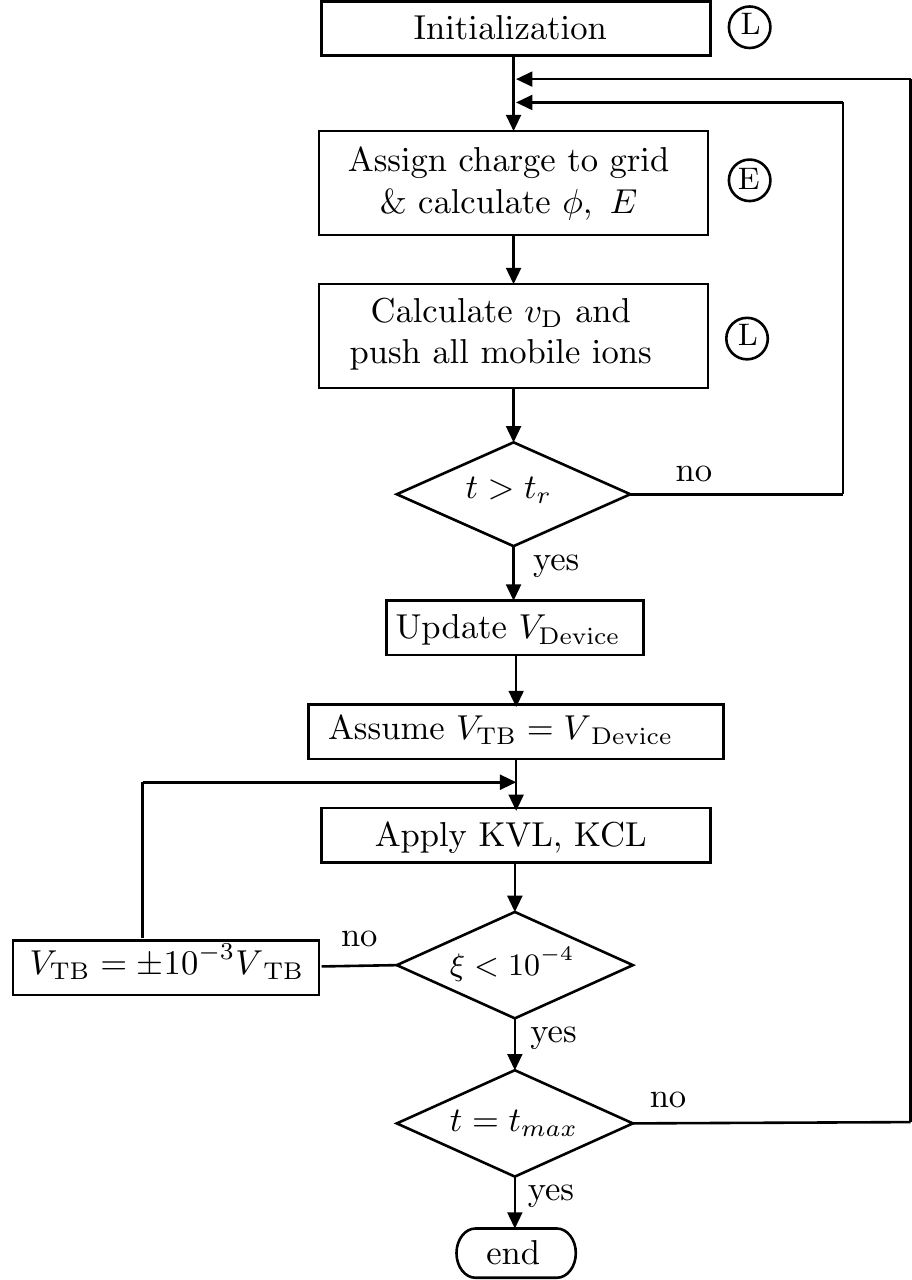}
\caption{Schematic flow diagram of the simulation approach. L represents the Lagrangian grid, and E represents the Eulerian grid.
$\xi$, $t_{r}$ and $t_{max}$ are the relative error between currents, relaxation time and the maximum simulation time, respectively.}
\label{fig:3-1} 
\end{figure}

\noindent Here, $\bar{d}$ is the average distance between the mobile ions and  Schottky contact, and $\bar{d}_{\rm r}$ is the initial average distance. $\bar{x}_{i}$ is the position of $i^{\rm th}$ mobile ion, $\bar{x}_{\rm SC}$ is the position of Schottky contact and $N_{\rm ions}$ is the number of mobile ions. The flowchart of the overall implementation of the simulation approach is shown in Fig.~\ref{fig:3-1}.

\subsection{Stochasticity modelling}
\label{sec:sec3.2} 
An essential property of the considered oxide-based DBMD is its intrinsic stochastic nature. The stochastic behaviour is usually measured in terms of cycle-to-cycle (C2C) variability and device-to-device (D2D) variability. It has been experimentally observed for a DBMD as shown in Figs.~\ref{fig:4-2}(a) and \ref{fig:4-5}(a). 
So far, the stochasticity in the model is only induced by random initial positioning of the mobile and fixed ions in the solid-state electrolyte.  However, this is insufficient to replicate the stochastic behaviour observed in a physical device since it only mimics the D2D variability. As the drift theory-based ion transport following Eq.~\eqref{Eq:3.1.5} is deterministic, the internal system state $q(t)$ also tends to become deterministic. However, $q(t)$ changes randomly as the mobile ions move randomly across the solid-state electrolyte for a physical device. So to account for stochasticity, we artificially perturb the internal state $q(t)$ of the device~\cite{Naous2016}. Notably, the internal state is changed indirectly by randomly perturbing the position of the mobile ions at each iteration of ion motion. So the position of the $i^{th}$ ion, $\bar{x}_{i(\rm s)}$ is given by

\begin{equation}
    \underset{\rm stochastic}{\underbrace{\bar{x}_{i(\rm s)}}} = \underset{\rm deterministic}{\underbrace{\bar{x}_{i(\rm d)}}} +  \hspace{5pt}\underset{\rm stochastic}{\underbrace{(r-0.5)\delta \bar{x}_{i(\rm d)}}},
    \label{Eq:3.1.8}
\end{equation}

\noindent where $r$ is a uniform (pseudo) random number between 0 and 1,  and $\delta$ denotes the maximum random displacement. The latter is chosen so that the underlying physical processes do not change and remain stable (i.e., $1\%- 5\%$ of $\bar{x}_{i(\rm d)}$). By modifying the position of the mobile ions, the average distance $\bar{d}$ also changes, which in turn makes $q(t)$ stochastic. Thus, the C2C variability feature of the memristive devices is mimicked.

\begin{figure*}[!t]
    \centering
    \begin{subfigure}
    {\includegraphics[width=0.98\textwidth]{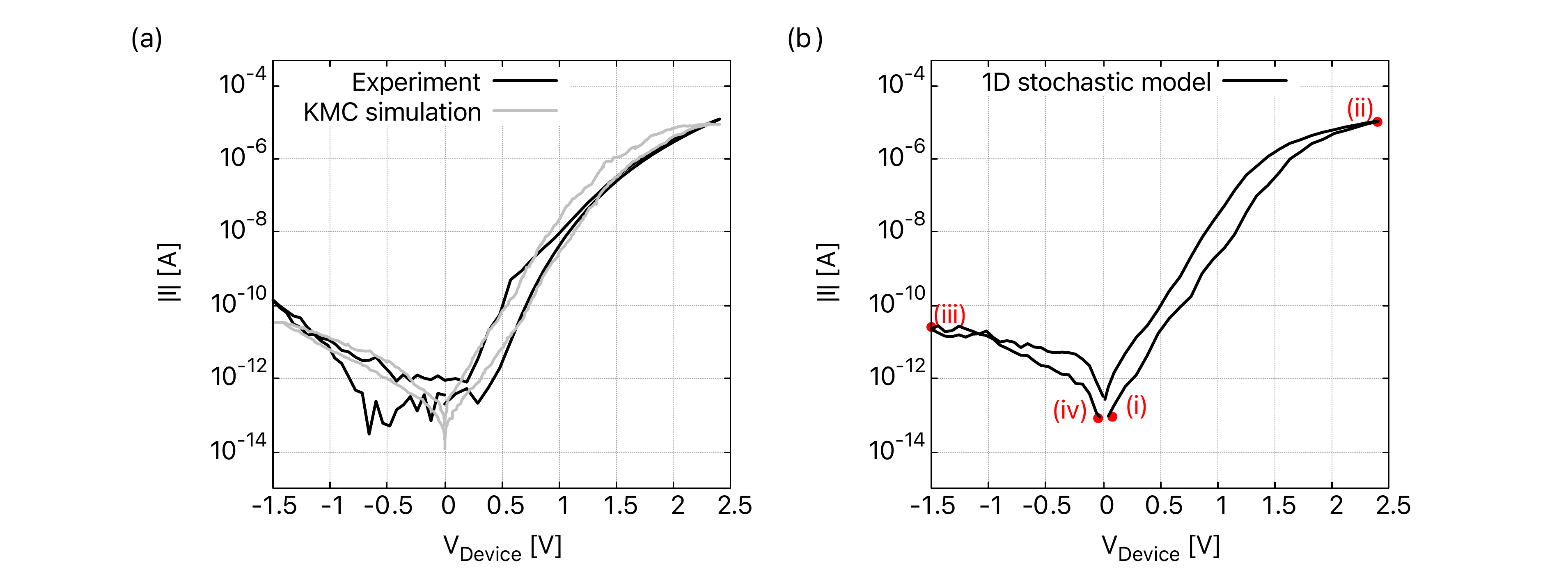}}
    \label{fig:4-1a}
    \end{subfigure}
    \begin{subfigure}
    {\includegraphics[width=0.95\textwidth]{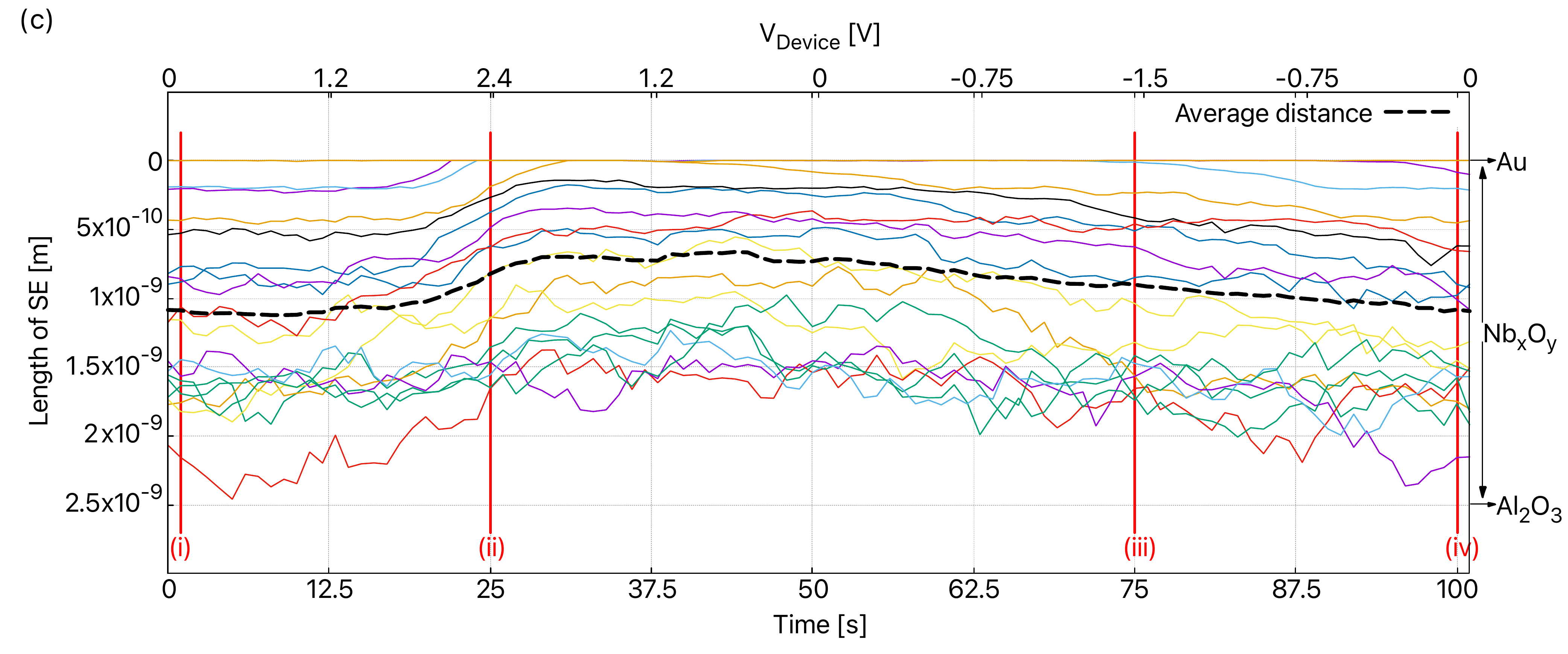}}
    \label{fig:4-1b}
    \end{subfigure}
    \caption{(a) Experimentally measured and kMC simulated current-voltage characteristics of the double barrier memristive device. (b) Calculated current-voltage characteristics of the double barrier memristive device obtained using the 1D stochastic model. (c) The positions of the mobile negatively charged ions as an outcome of the simulation are shown for different instants of time and voltage. The ion movement is shown for a single applied voltage cycle. The coloured lines represent twenty different mobile ions, and the black dashed line represents the absolute average distance ($\bar{d}$). The red lines indicate the corresponding positions marked in subfigure~(b).}
    \label{fig:4-1}
\end{figure*}


\subsection{\label{sec:sec3.3} Implementation of current mechanisms}

The calculation of current through the different regions of the DBMD is explained in the following sections:

\subsubsection{Tunnelling current}
\label{sec:sec-3.3.1}
The electronic current flow through a metal-insulator-metal (MIM) system can be given by the electric tunnel effect. The tunnelling phenomenon of electrons through the ${\rm Al_{2}O_{3}}$ tunnel barrier can be expressed using a set of equations derived by Simmons \cite{Simmons1963}. The general Simmons current equation is given as

\begin{equation}
    \begin{split}
    I_{\rm TB} &= \frac{eA_{\rm d}}{2\pi h\left (\beta d_{\rm eff}  \right )^{2}}\Biggl( \Phi_{\rm eff}^{\rm t} \cdot {\rm exp}\left \{- {\rm A}\sqrt{\Phi_{\rm eff}^{\rm t}}\right \} - \\
    &\left (\Phi_{\rm eff}^{\rm t}+ e\left | V_{\rm TB} \right |  \right ) \cdot {\rm exp}\left \{- {\rm A} \sqrt{\Phi_{\rm eff}^{\rm t}+ e\left | V_{\rm TB} \right | }\right \}\Biggr),
    \label{eq:3.3.1}
    \end{split}
\end{equation}

\noindent where

\begin{equation*}
    {\rm A} = \frac{4\pi\beta d_{\rm eff} \sqrt{2m} }{h}.
    \label{eq:3.3.2}
\end{equation*}

\noindent $\Phi_{\rm eff}^{\rm t}$  is the effective tunnel barrier height, $d_{\rm eff}$ is the effective tunnel barrier width, $A_{\rm d}$ is the device area, and $\beta$ is a correction factor. $e$, $m$, and $h$ are the elementary charge, the free electron mass, and the Planck constant, respectively. The absolute value of $V_{\rm TB}$ is considered here, assuming that Eq.~\eqref{eq:3.3.1} is symmetric between
positive and negative voltage biases. Since the Simmons formula is only considered for elastic tunnelling, the resulting electronic current mainly depends on the local ion concentration at the $\rm Al_{2}O_{3}/Nb_{x}O_{y}$ contact. So, the effective tunnel barrier width and the effective tunnel barrier height are used here, which are given by

\begin{equation}
    d_{\rm eff} = d_{\rm 0}(1 + \lambda_{\rm d}\,q(t))
    \label{eq:3.3.3}
\end{equation}

\noindent and 

\begin{equation}
    \Phi_{\rm eff}^{\rm t} = \Phi^{\rm t}(1 + \lambda_{\rm t}\,q(t)).
    \label{eq:3.3.11}
\end{equation}

\noindent $d_{\rm 0}$ is the actual width of the tunnel barrier, $\Phi^{\rm t}$ is the actual tunnel barrier height. $\lambda_{\rm d}$ and $\lambda_{\rm t}$ are the fitting parameters.
The fitting parameters are chosen in a way that the simulation outcome leads to specific  experimental behaviour, and their values can be between 0 and 1.

\subsubsection{Ohmic current}
\label{sec:sec-3.3.2}

The current across the ${\rm Nb_{x}O_{y}}$ solid-state electrolyte ($I_{\rm SE}$) can be calculated directly using Ohm's law as,

\begin{equation}
    I_{\rm SE} = \sigma A_{\rm d}\frac{V_{\rm SE}}{l_{\rm SE}},
    \label{eq:3.3.4}
\end{equation}

\noindent where $\sigma$ is the conductivity of ${\rm Nb_{x}O_{y}}$ and $l_{\rm SE}$ is the length of ${\rm Nb_{x}O_{y}}$. Also, from Eq.~\eqref{Eq:2.2}, $I_{\rm SE} = I_{\rm TB}$, so

\begin{equation}
    V_{\rm SE} = I_{\rm TB}\frac{l_{\rm SE}}{\sigma}.
    \label{eq:3.3.5}
\end{equation}


\subsubsection{Schottky current}
\label{sec:sec-3.3.3}

In the presence of a high electric field, the oxygen ions move close to the $\rm Nb_{x}O_{y}$/Au Schottky interface and back to their inertial position. The concentration of these ions at the Schottky contact considerably affects the Schottky barrier height and the ideality factor. The current conduction mechanism across this Schottky contact is described by thermionic emission theory~\cite{Sze2007}. According to this theory, the Schottky diode current is given by

\begin{equation}
    I_{\rm SC} = I_{\rm R}\left ( {\rm exp}\left \{ \frac{eV_{\rm SC}}{n_{\rm eff}k_{B}T} \right \} - 1\right ),
    \label{eq:3.3.6}
\end{equation}

\noindent where $n_{\rm eff}$ is the effective ideality factor which describes the deviation from an ideal current, $k_{B}$ is the Boltzmann constant, and $T$ is the temperature. The reverse current, $I_{\rm R}$ for different voltage polarities, is given by

\textup{forward bias: }
\begin{equation}
    I_{\rm R, V_{SC}> 0} = A_{\rm d}A^{*} T^{2}{\rm exp}\left \{ \frac{-\Phi_{\rm eff}^{b}}{k_{B}T} \right \},
    \label{eq:3.3.7}
\end{equation}

\textup{reverse bias:}
\begin{equation}
        I_{\rm R, V_{SC}<  0} = A_{\rm d}A^{*} T^{2}{\rm exp}\left \{ \frac{-\Phi_{\rm eff}^{b}}{k_{B}T} \right \}
        {\rm exp}\left \{ \frac{\alpha _{r}\sqrt{\left | V_{\rm SC} \right |}}{k_{B}T} \right \}.
        \label{eq:3.3.8}
\end{equation}

\begin{figure}[!t]
    {\includegraphics[width=1.0\columnwidth]{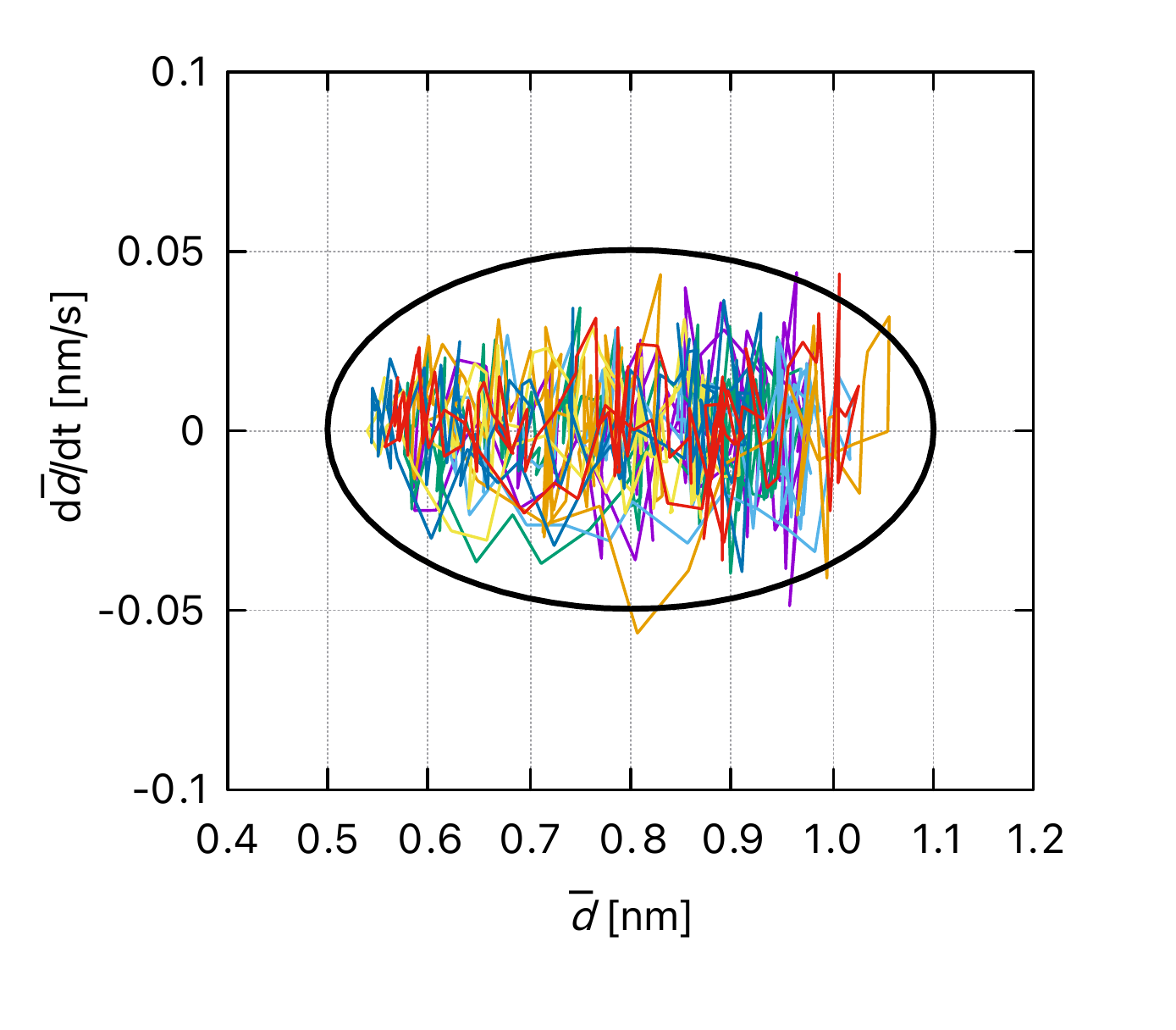}}
    \caption{The phase-space plot of the average absolute distance, $\bar{d}$. The plot is obtained for ten consecutive cycles of the applied voltage, shown in distinct colours. It is a Poincar$\acute{\rm e}$ plot describing the periodicity of the stochastic model.}
    \label{fig:4-4}
\end{figure}

\noindent Here $\Phi_{\rm eff}^{b}$ is the effective Schottky barrier height, and $A^{*} \rm{=  1.20173 \times 10^{6} A/ (m^{2}K^{2})}$ is the effective Richardson constant. $\alpha_{r}$ is a device-dependent parameter that describes the voltage dependency of the reverse current as measured experimentally. The reverse current during the set process is dominated by lowering the Schottky barrier height, decreasing gradually with the applied positive bias. While the Schottky barrier height increases during the reset process. By considering the spatial rearrangement of mobile ions due to the high electric field, the effective Schottky barrier height and effective ideality factor are defined as 

\begin{equation}
    \Phi_{\rm eff}^{b} = \Phi^{\rm b0}(1 + \lambda_{\rm b}\, q(t)),
    \label{eq:3.3.9}
\end{equation}

\begin{equation}
    n_{\rm eff} = n_{\rm 0}(1 + \lambda_{n}\,q(t)).
    \label{eq:3.3.10}
\end{equation}

\noindent $\Phi^{\rm b0}$ is the initial Schottky barrier height, and $n_{0}$ is the initial ideality factor.$\lambda_{\rm b}$ and $\lambda_{\rm n}$ are the fitting parameters.

\begin{figure*}[!t]
    \centering
    \begin{subfigure}[]
    {\includegraphics[width=1.0\columnwidth]{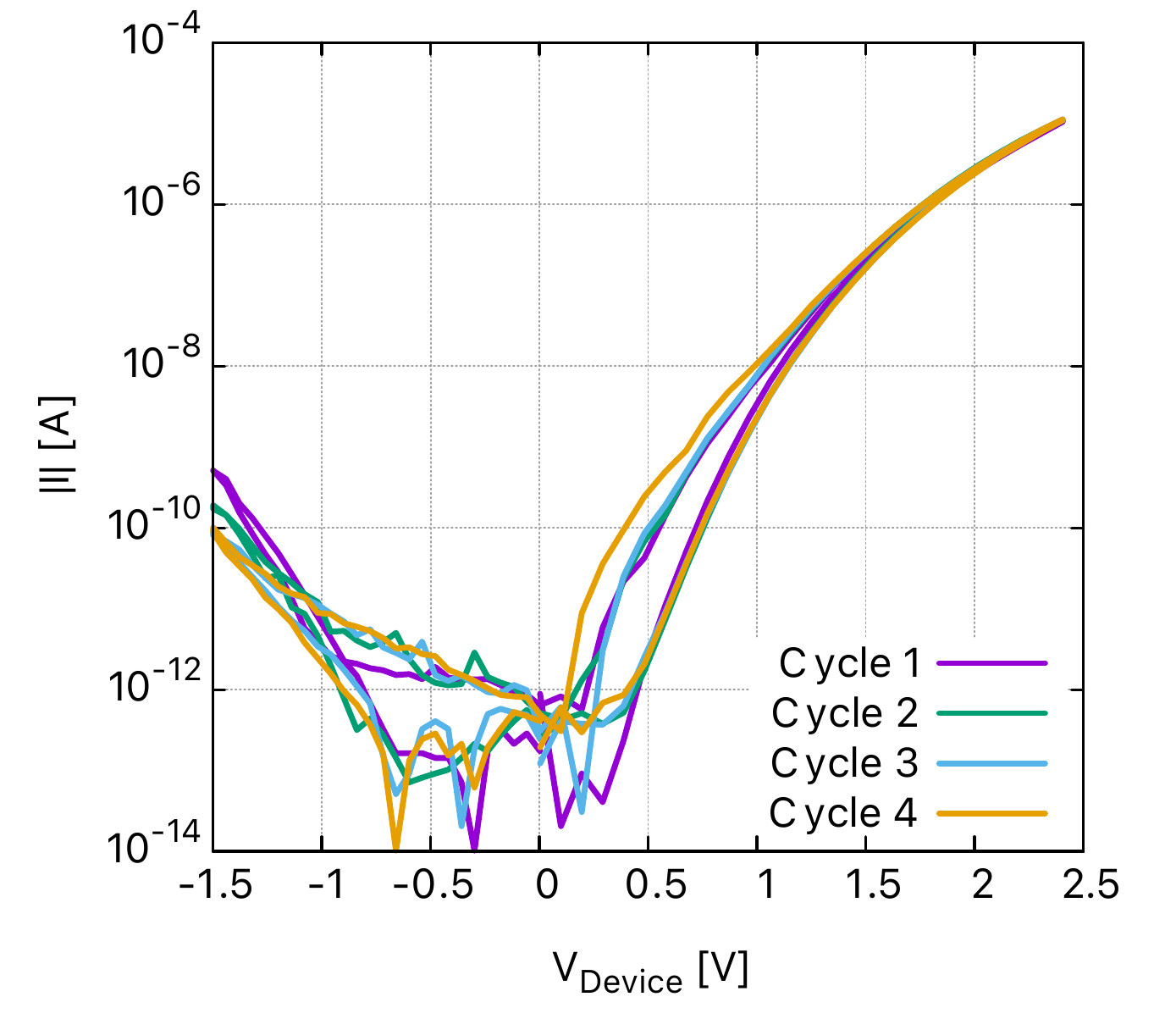}}
    \label{fig:4-2a}
    \end{subfigure}
    \begin{subfigure}[]
    {\includegraphics[width=1.0\columnwidth]{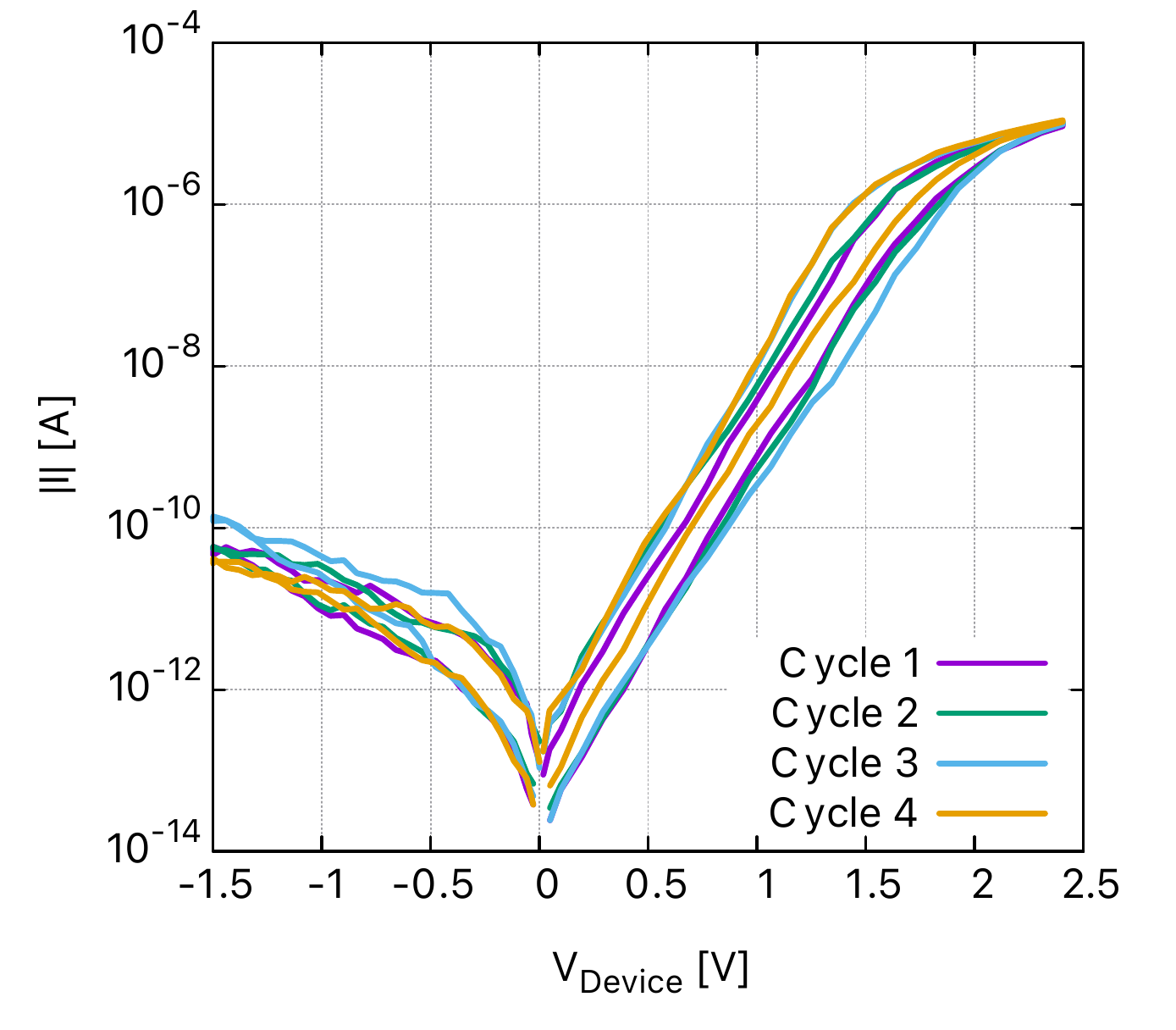}}
    \label{fig:4-2b}
    \end{subfigure}
    \caption{\textit{I}-\textit{V} curves of DBMD showing the C2C variability obtained for an applied voltage of four consecutive cycles. (a) Experimental results and (b) simulation results.}
    \label{fig:4-2}
\end{figure*}

\begin{figure*}[!t]
    \centering
    \begin{subfigure}[]
    {\includegraphics[width=1.0\columnwidth]{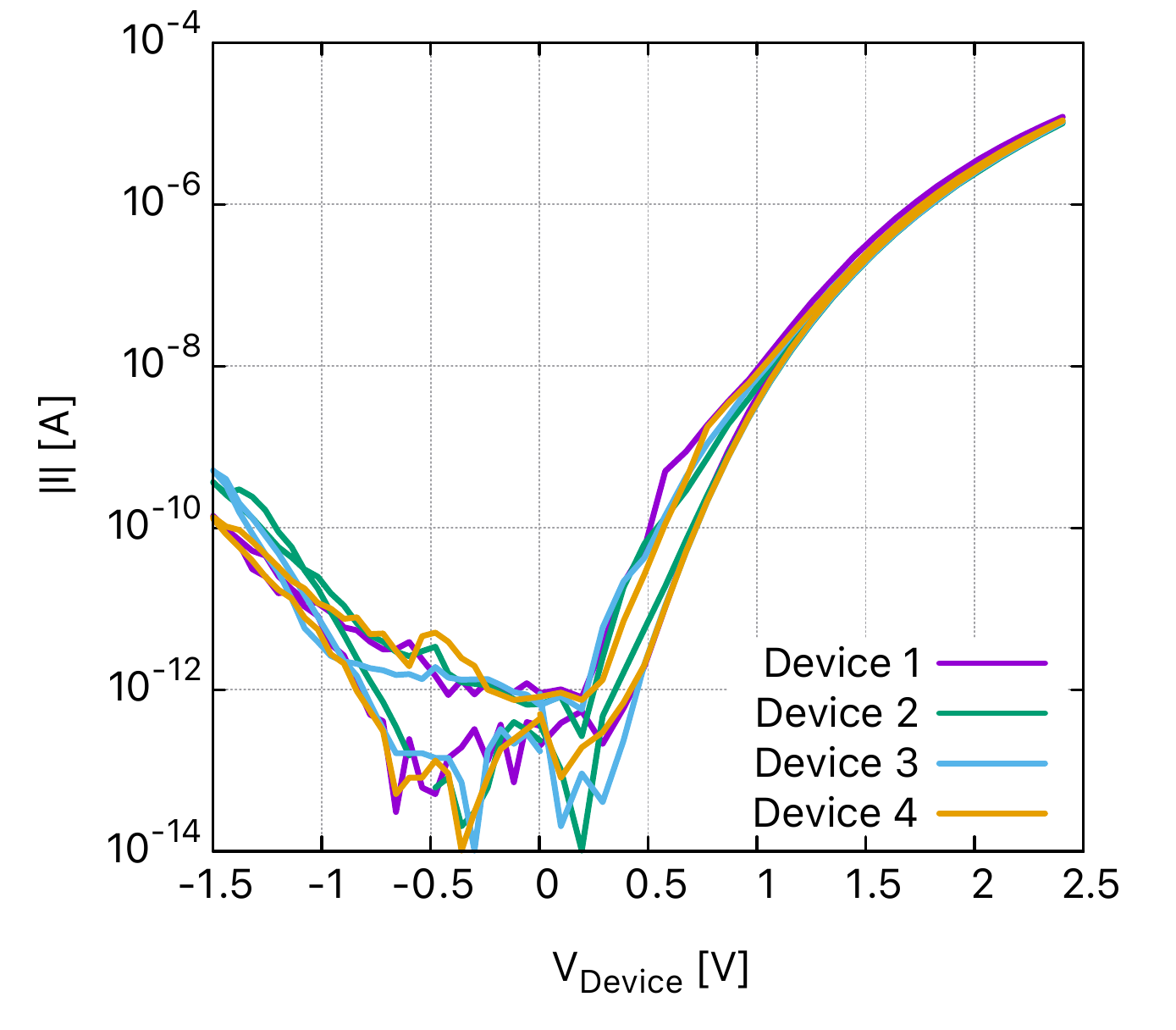}}
    \label{fig:4-5a}
    \end{subfigure}
    \begin{subfigure}[]
    {\includegraphics[width=1.0\columnwidth]{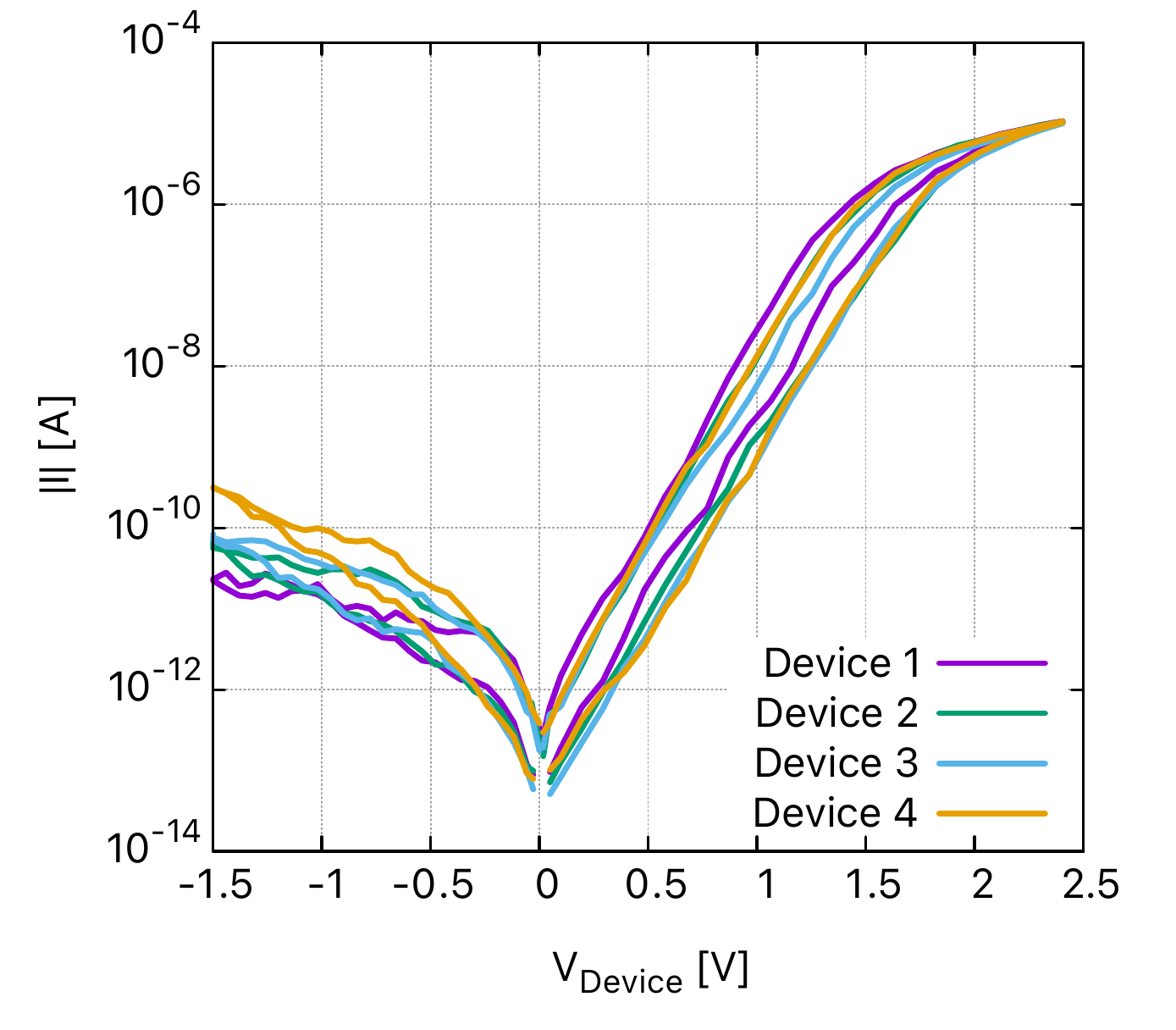}}
    \label{fig:4-5b}
    \end{subfigure}
    \caption{\textit{I}-\textit{V} curves of DBMD showing the D2D variability obtained for a single applied voltage cycle, but with four different initial ion arrangement. (a) Experimental results and (b) simulation results.}
    \label{fig:4-5}
\end{figure*}

\section{\label{sec:sec4.0} Results and discussion}

The current-voltage characteristics (\textit{I}-\textit{V} curve) of the double barrier memristive device calculated using the proposed stochastic model is compared to the \textit{I}-\textit{V} curve measured experimentally~\cite{Hansen2015} and the \textit{I}-\textit{V} curve calculated using the kMC simulations~\cite{Dirkmann2016} in Figs.~\ref{fig:4-1} for a single cycle. For the simulation, a linear voltage sweep was applied from 0V to 3V and back to 0V, in order to set the device from HRS to LRS. To reset the device to its initial HRS, the voltage was ramped linearly from 3V to -1.5V and then back to 0V. Then the resultant \textit{I}-\textit{V} curve, shown in Fig.~\ref{fig:4-1}(a), resembles a typical memristive hysteresis loop. As observed, the \textit{I}-\textit{V} curve extracted from the 1D stochastic model (shown in Fig.~\ref{fig:4-1}(b)) is qualitatively in good agreement with the measured \textit{I}-\textit{V} curve and also the kMC simulated \textit{I}-\textit{V} curve, but quantitatively the three \textit{I}-\textit{V} curves show slight variation. However, this minute variation is acceptable considering that the device and the simulation models are stochastic.

To investigate the movement of mobile ions in the solid electrolyte under the influence of an applied voltage and their influence on the device resistance, the Spatio-temporal plot of ion transport for a single applied voltage cycle is shown in Fig.~\ref{fig:4-1}(c). The coloured lines in the plot represent 20 different ion movements that are distributed randomly across the solid-state electrolyte. The absolute average position of the ions, $\bar{d}$ displayed as a black dashed line in Fig.~\ref{fig:4-1}(c), indicates the internal state of the device. Additionally, the four significant positions highlighted in red correspond to analogous points on the \textit{I}-\textit{V} curve in Fig.~\ref{fig:4-1}(b). Initially, at point (i), all the ions are randomly scattered across the simulation domain (corresponding to $\bar{d}=1.1$\,nm). When a positive voltage is applied, the ions start to drift towards the Schottky interface (indicated as point (ii)), thus changing the device state from HRS to LRS. For a negative voltage ramp, the ions drift in the opposite direction, and the value of $\bar{d}$ is then close to the initial value at -1.5V (point (iii)). The voltage is then increased from -1.5V to 0V to reset the device completely (point (iv)).

Since the process is stochastic, the value of $\bar{d}$ is variable but should be within a range of approximately $\pm \delta$ of $\bar{d}$\,(deterministic). This can be evaluated using the phase-space plot of $\bar{d}$ and the derivative of $\bar{d}$ with respect to the time, as shown in Fig.~\ref{fig:4-4}. The plot is obtained for the movement of ions recorded for ten consecutive applied voltage cycles (each colour corresponds to a single cycle of the applied voltage). It is a kind of Poincar$\acute{\rm e}$ plot that describes the periodic behaviour of the model. Although the trajectories of $\bar{d}$ in the phase-space plot are distinct and random for all applied voltage cycles, their ensemble is still within an arbitrary range of $\pm 5\%$ of $\bar{d}$\ (black ellipse shown in Fig.~\ref{fig:4-4}), hence proving the periodicity and stochasticity of the proposed model. The range determined by the black ellipse depends on the value of $\delta$, which is taken as $5\%$ in the manuscript.

The stochasticity of the proposed model can be further determined by observing the \textit{I}-\textit{V} curves in Fig.~\ref{fig:4-2} and \ref{fig:4-5}. The plots show the simulated and measured C2C variability and D2D variability of DBMD obtained under comparable conditions. Firstly, to check for the C2C variability, the simulation was performed for four consecutive cycles of the applied voltage.  As a result, four different \textit{I}-\textit{V} curves are observed in Fig.~\ref{fig:4-2}(b), which follow the same trend yet have slightly varying $\rm R_{on}/ R_{off}$ ratios. The deviation from one cycle to another can be reduced or increased by considering a smaller or larger value for $\delta$, respectively. Similarly, to verify the D2D variability, the simulation is performed four times with different initial ion arrangements. The difference in the ion arrangements here corresponds to having four different real-world DBMDs. The resultant \textit{I}-\textit{V} curves in Fig.~\ref{fig:4-5}(b) are distinct from each other with different $\rm R_{on}/ R_{off}$ ratios. Thus proving the stochasticity of the proposed model, which very well reproduces the inherent stochastic behaviour of real-world DBMDs shown in Figs.~\ref{fig:4-2}(a) and \ref{fig:4-5}(a).  

In comparison to the state of the art simulation models, the proposed model shows a significant C2C and D2D variation, similar to experimental results. This makes the proposed model well suited for performing circuit simulations in artificial intelligence (AI) computing, reconfigurable logic computing, or hardware security primitives, which mainly rely on the stochastic behaviour of memristive devices. Moreover, the set current is nearly the same for experimental and calculated \textit{I}-\textit{V} curves in both C2C and D2D plots (see Figs.\,\ref{fig:4-2} and \ref{fig:4-5}). This is because, during the set process (LRS), most of the ions are present at the Au interface, regardless of their initial position and applied voltage cycle. However, due to their random movement, the ions do not migrate back to their original position during the reset process. In other words, the value of $\bar{d}$ for LRS is relatively similar for all cycles, but $\bar{d}$ is always different for HRS. Therefore, the reset current, which depends on $\bar{d}$, is always different. This is clearly noticeable in both experimental and calculated \textit{I}-\textit{V} curves. 

\begin{figure}[!t]
    {\includegraphics[width=1.0\columnwidth]{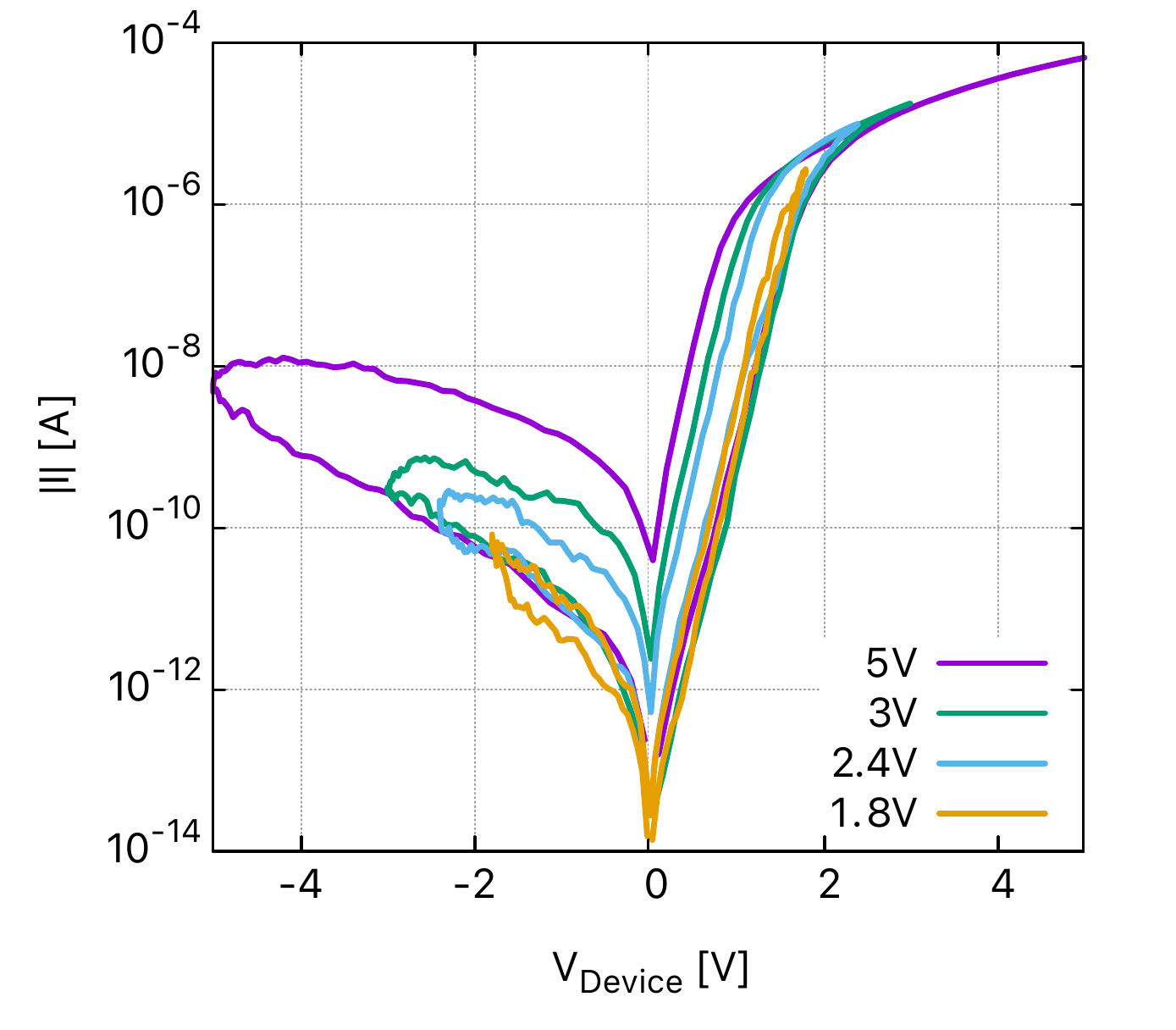}}
    \caption{The current-voltage characteristics of DBMD obtained for a sinusoidal applied voltage, showing the effect of maximum applied voltage on the hysteresis.}
    \label{fig:4-3}
\end{figure}

The applied voltage used to obtain the plots in Fig.~\ref{fig:4-1} follows an unsymmetrical triangular waveform, particularly a sawtooth waveform. The sawtooth waveform was chosen solely to study the set and reset process in a device. Using the proposed 1D stochastic model, the response of the device to other input signals with different set and reset voltages can also be studied. Fig.~\ref{fig:4-3} shows the \textit{I}-\textit{V} curve of a DBMD obtained using the proposed model for a sinusoidal applied voltage. The \textit{I}-\textit{V} curves are obtained for four different maximum applied voltages (2V, 2.5V, 3V and 5V). As expected, the \textit{I}-\textit{V} curves show smooth transitions at the maximum positive and negative bias. However, for the positive branch of the hysteresis, asymptotic behaviour is observed for voltages more than 3V. Since the set voltage for DBMD is 3V (as mentioned in \cite{Hansen2015}), the device does not show any significant change for voltages more than 3V. Moreover, it is also verified experimentally that the width of the hysteresis increases as the maximum applied voltage increases, whereas hysteresis effects collapse at lower voltages~\cite{Dirkmann2016,Maestro2021}. The physical devices suffer a dielectric breakdown at high voltages around $\pm\,5$V and even for lower absolute values of negative voltages.

\section{\label{sec:sec5.0} Conclusion}
A circuit simulator compatible 1D stochastic model for simulating interface type memristive devices (such as DBMD, BFO) is proposed in this paper.  The model integrates the stochastic behaviour of a real-world DBMD with a realistic ion transport by consistently coupling Newton's laws with a Poisson solver, inspired by the Cloud-In-a-Cell scheme. The DBMD current-voltage characteristics obtained using the proposed model are in good agreement with experimental results. Furthermore, different hysteresis relations are obtained while performing several simulations, describing the stochastic nature of the model. Despite being a straightforward 1D model with minimum mathematical formulations, the proposed model successfully reproduced all the essential features of a real-world DBMD. These features include the non-linear analogue behaviour, intrinsic stochastic behaviour, an asymmetry between positive and negative bias, and high voltage current saturation. Therefore, the proposed stochastic model is perceived as a powerful tool to simulate large-scale memristive circuits in areas such as bio-inspired neural networks, hardware security or electrical networks. Additionally, the model could be used to study the dynamics of other non-filamentary memristive devices with minimal computational resources.

\begin{acknowledgments}
Funded by the Deutsche Forschungsgemeinschaft (DFG, German Research
Foundation) - Project-ID 434434223 - SFB 1461 and Project-ID 439700144
- Research Grant MU 2332/10-1 in the frame of Priority Program SPP 2253.
\end{acknowledgments}

\section*{ORCID IDs}

\noindent S. Yarragolla: \href{https://orcid.org/0000-0002-2973-4943}{https://orcid.org/0000-0002-2973-4943} \\
T. Hemke: \href{https://orcid.org/0000-0003-2436-5840}{https://orcid.org/0000-0003-2436-5840}\\
J. Trieschmann: \href{https://orcid.org/0000-0001-9136-8019}{https://orcid.org/0000-0001-9136-8019}\\
H. Kohlstedt: \href{https://orcid.org/0000-0002-5181-8633}{https://orcid.org/0000-0002-5181-8633}\\
\noindent T. Mussenbrock: \href{https://orcid.org/0000-0001-6445-4990}{https://orcid.org/0000-0001-6445-4990}

\bibliography{bibliography}

\end{document}